\documentclass[preprint,12pt,sort&compress]{elsarticle}
\usepackage{multirow}
\usepackage{overpic}
\usepackage{color}

\usepackage{amssymb}

\journal{Nucl. Instr. and Meth. A}

\begin{document}

\newcommand{\sigmaARM}{\sigma_{\mbox{\scriptsize ARM}}}
\newcommand{\Eabs}{E_{\mbox{\scriptsize abs}}}
\newcommand{\Escat}{E_{\mbox{\scriptsize scat}}}
\newcommand{\Etot}{E_{\mbox{\scriptsize tot}}}
\newcommand{\thetaC}{\theta_{\mbox{\scriptsize C}}}
\newcommand{\thetaR}{\theta^{\mbox{\scriptsize R}}}
\newcommand{\phiR}{\phi^{\mbox{\scriptsize R}}}
\newcommand{\thetaCR}{\theta_{\mbox{\scriptsize C}}^{\mbox{\scriptsize R}}}
\newcommand{\thetageom}{\theta_{\mbox{\scriptsize geom}}}

\begin{frontmatter}



\title{First demonstration of a Compton gamma imager based on silicon photomultipliers}


\author{P.R.B.~Saull$^{1}$, L.E.~Sinclair$^{2}$, H.C.J.~Seywerd$^{2}$, D.S.~Hanna$^{3}$, P.J.~Boyle$^{3}$, A.M.L.~MacLeod$^{3}$}

\address{$^{1}$Institute for National Measurement Standards, National Research Council, 1200~Montreal Rd, Ottawa, Ontario, K1A 0R6, Canada
\\$^{2}$Geological Survey of Canada, Natural Resources Canada, 601 Booth St, Ottawa, Ontario, K1A 0E8, Canada
\\$^{3}$Physics Department,McGill University, 3600 University St, Montreal, Quebec, H3A 2T8, Canada}

\begin{abstract}
We are developing a rugged and person-transportable Compton gamma imager for use in security investigations of radioactive materials, and for radiological incident remediation.
The imager is composed of layers of scintillator with light collection for the forward layers
provided by silicon photomultipliers and for the rear layer by photomultiplier tubes.
As a first step, we have developed a 1/5$^{\mbox{\scriptsize th}}$-scale demonstration unit of the final imager.
We present the imaging performance of this demonstration unit for 
$^{137}$Cs  at angles of up to 
30$^\circ$ off-axis.  Results are also presented for 
$^{113}$Sn  and $^{22}$Na.
This represents the first demonstration of the use of silicon photomultipliers as an embedded component for light collection in a Compton gamma imager.

\end{abstract}


\end{frontmatter}


\section{Introduction}
\label{sec:intro}

A Compton gamma imager can provide information about the location and distribution of gamma emitting radioactive sources in an intuitive graphic format that would be very useful both in security investigations and in radiological incident consequence management.

Compton gamma imaging relies on collection of the energies and positions of the near-simultaneous deposits caused by a gamma ray as it scatters in a segmented detector.  Typically, the first few layers of the detector are used to initiate a Compton scatter with energy deposit $E_1$, and are referred to as the scatter detector.  The final layer collects the energy deposited by the scattered gamma ray, $E_2$, and is called the absorber detector.  For events with just these two energy deposits, the Compton scatter angle, $\thetaC$,
 can be reconstructed according to,
\begin{equation}
   \thetaC = 1 + m_e c^2 (1/\Etot - 1/E_2),
\end{equation}
where $\Etot = E_1 + E_2$, $m_e$ is the electron rest mass and $c$ is the speed of light.
Thus, the original source position can be reconstructed up to an arbitrary azimuthal angle, giving a cone-shaped locus on which the source should lie.  By overlaying several of these ``Compton cones'' a picture of the emitter and its location may be obtained~\cite{Nature_1974}.  Note that a Compton imager is also inherently spectroscopic, giving it the ability to identify the isotope if the source is a priori unknown.

Any scattering in dead material within the scatter detector or between the scatter and absorber detectors will lead to inefficiencies, or worsening of image resolution.  Therefore, low-density components have typically been used in the scatter planes, such as gaseous tracking chambers~\cite{TPC_Japan_2005} or silicon gamma detectors~\cite{Si_Japan_2009, Si_Naval_2007, Si_LANL_2006, Si_LLNL_2006, Si_Michigan_1998}.

Nevertheless, the capabilities of solid scintillator in terms of energy resolution, high efficiency, ruggedness and affordability has warranted its investigation for use in the scatter detector~\cite{us_2009, them_2011}.  Compton gamma imagers have been demonstrated which employ inorganic scintillator in the scatter layer with either conventional photomultiplier tube (PMT) readout in the path of the incoming gamma ray~\cite{SORDS_2009} or with a novel scintillator bar concept such that the PMTs can be located at the side of the detector, out of the gamma-ray 
path~\cite{us_IEEE_2011}.

Recently, it has been shown that silicon photomultipliers (SiPMs), which are essentially arrays of many tiny Geiger-mode avalanche photodiodes, can be used as light collection devices for scintillators, yielding excellent timing characteristics and energy 
resolutions~\cite{SiPM_review_2011,SiPM_rev_2006,SiPM_rev_2004}.

We are taking advantage of the low mass and small form factor of the SiPM to develop a Compton gamma imager using layers of scintillator in which the scatter layers are read out by embedded SiPMs.  As extra mass at the rear of the absorber detector does not interfere with the events of interest, the absorber layer is read out by PMTs.  Here, we present the performance characteristics of a 1/5$^{\mbox{\scriptsize th}}$-scale version of this Compton gamma imager, for the energies 390~keV, 662~keV and 1274~keV, as provided by the isotopes
$^{113}$Sn, $^{137}$Cs and $^{22}$Na, respectively.  The performance is characterized in terms of energy resolution, angular resolution measure, efficiency, and time to image.

\section{Experimental Setup}

\label{sec:setup}
\begin{figure}
   \begin{center}
   \begin{tabular}{c}
   \begin{overpic}[height=4.9cm]{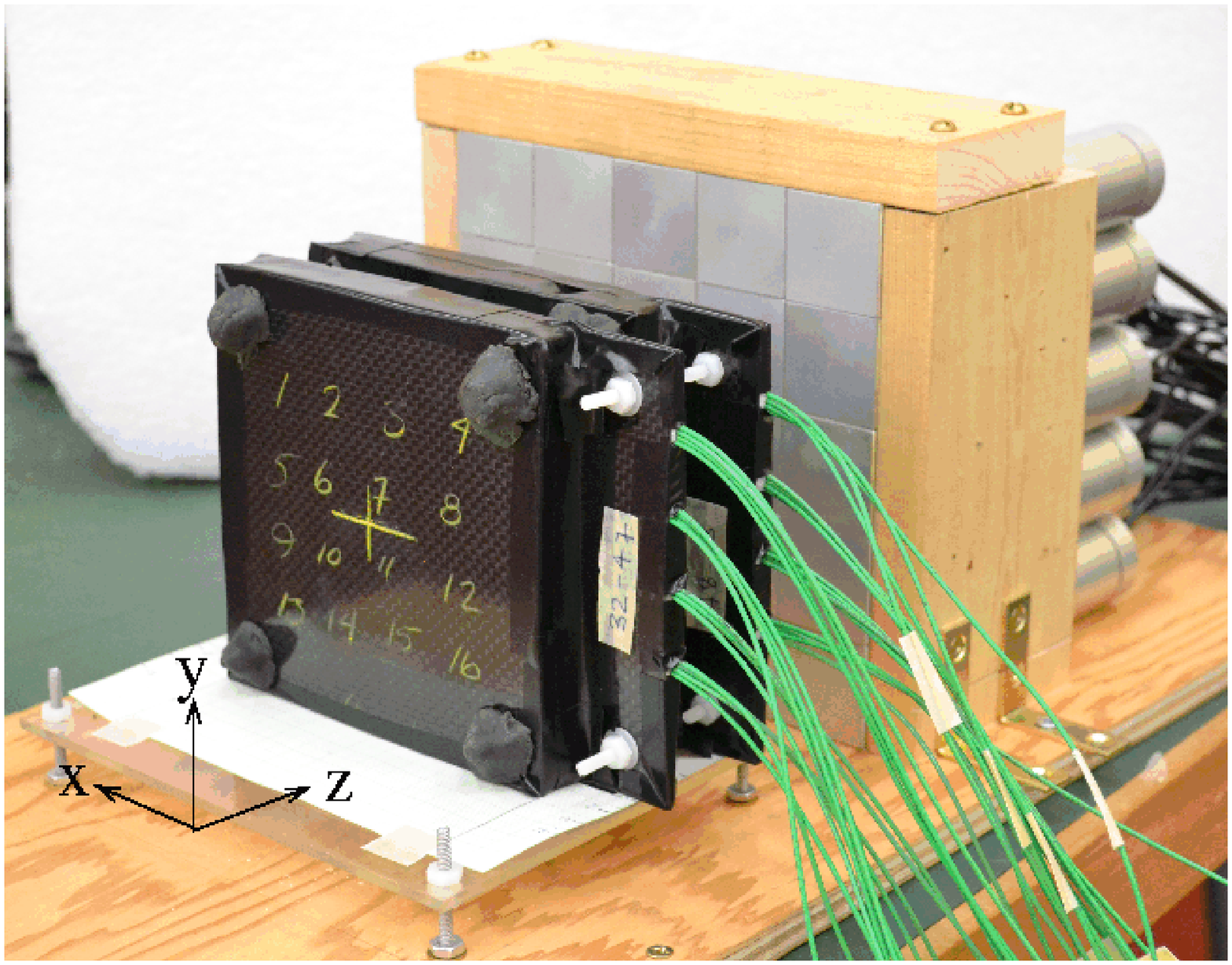}\put(8,67){a)}\end{overpic}
   \hspace{0.4cm}
   \begin{overpic}[height=4.9cm]{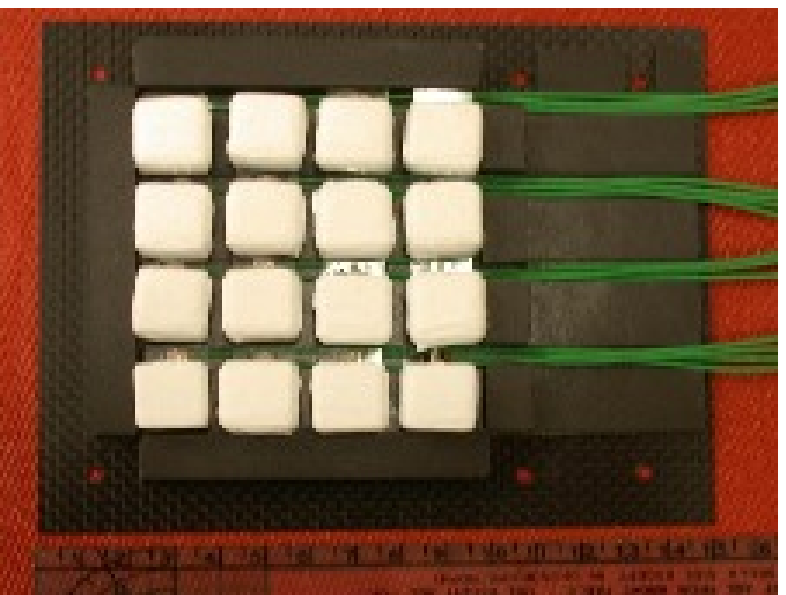}\put(8,65){\textcolor{white}{b)}}\end{overpic}\\
    \\
    \begin{overpic}[height=4.9cm]{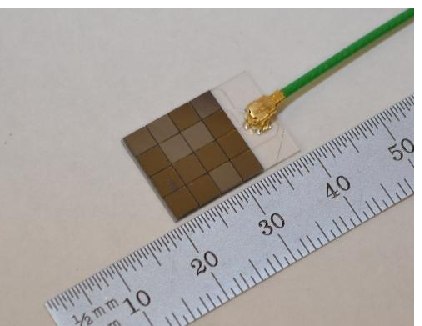}\put(8,65){\textcolor{white}{c)}}\end{overpic}
   \hspace{.4cm}
   \begin{overpic}[height=4.9cm]{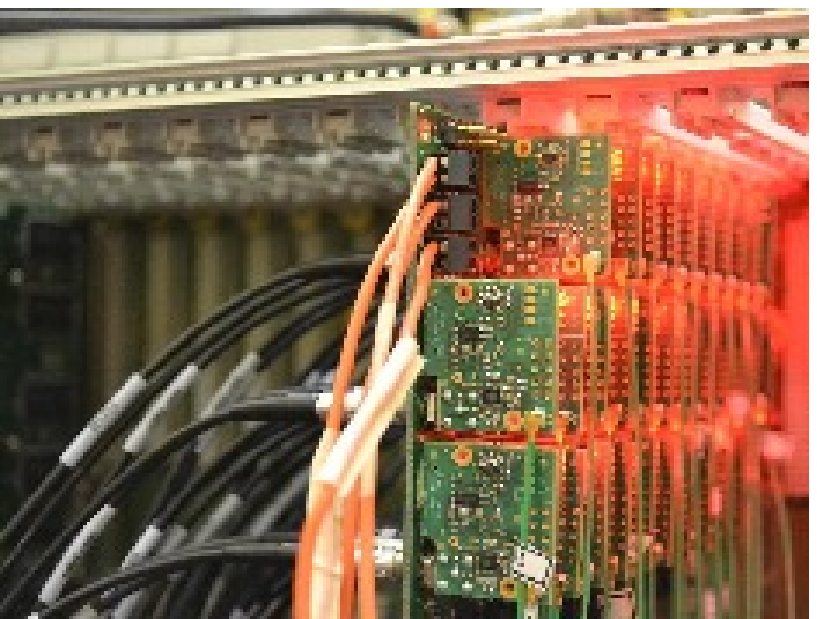}\put(8,65){\textcolor{black}{d)}}\end{overpic}
   \end{tabular}
   \end{center}
   \caption
   { \label{fig:diagram} 
a) The Compton gamma imager including the first and second scatter layers, and the absorber layer.  b) The inside of a scatter layer, without the foam alignment spacers.  The sixteen CsI(Tl) crystals are shown wrapped with plumber's tape.  The glass of the SiPM is visible, as is the micro coax cable.  c) A single SiPM detector unit.  d) The pre-amplification boards for the thirty-two SiPM+CsI(Tl) channels, connected to the motherboard.}
   \end{figure}

Our setup, and various components of the detector, are shown in Figure~\ref{fig:diagram}.
Fig.~\ref{fig:diagram}~a) shows the imager as a whole.  The two scatter layers are on the left, in the front, and the absorber layer is on the right, 
at the rear.  
The scatter layers are parallel to the absorber layer and centred on it.  The centre of the forward scatter layer is situated 10.5~cm in front of the front face of the absorber layer.  The centre of the rear scatter layer is situated 7~cm in front of the front face of the absorber layer.

The coordinate system has its origin at the centre of the front face of the front scatter layer.  The z-axis is the axis of symmetry of the imager, with the y axis pointing up, and the x-axis horizontal, as indicated in 
Fig.~\ref{fig:diagram}~a).

The two scatter layers each consist of four-by-four arrays of (1.35 cm)$^3$ cubes of CsI(Tl) scintillating crystals read out by SiPMs.  The CsI(Tl) crystals are mated to the SiPMs using NyoGel OC-462 optical 
gel\footnote{Nye Lubricants Inc, Fairhaven, MA, USA}.  The CsI(Tl) crystal and SiPM are wrapped together with several layers of white plumber's tape.  
The CsI(Tl) plus SiPM ``pixels'' are assembled on a 1/32~inch-thick carbon fibre sheet, with adhesive foam used to cushion the SiPMs and to position the individual pixels with a grid spacing of 2~cm.
A second carbon fibre layer, with foam padding, is bolted to the top of the CsI(Tl) crystals to sandwich them together with the SiPMs.
The exterior dimension of each carbon fibre sandwich in x and y is 12~cm x 12~cm (excluding a cable clamp).  The sandwiches are 2~cm thick.
An open scatter layer is shown in Fig.~\ref{fig:diagram}~b) where the foam spacers in between the pixels have been left out in order to display the individual pixels and show the glass and cable connector of the SiPM light collectors.

The SiPM detector units were custom manufactured for this work by 
SensL\footnote{SensL Technologies Ltd, Cork, Ireland}.
They were designed to have a very small footprint and to present very little dead material in the path of the incoming gamma rays.
The dimension of each SiPM detector unit is 13.32~x~19.00~x~0.98~mm$^3$.  Each consists of a four by four matrix of 3~x~3~mm$^2$ die size silicon photomultipliers, on a glass substrate.  There are a total of 76~384 Geiger-mode avalanche photodiodes in each SiPM detector unit.
An SiPM detector unit is shown in Fig.~\ref{fig:diagram}~c).
Herein the entire SiPM detector unit is referred to simply as an SiPM.  
Bias supply and signal readout for the SiPM is accomplished via micro-coaxial cable, also shown in Fig.~\ref{fig:diagram}~c).

SensL custom front-end boards provide both the SiPM bias voltage supply and pre-amplification.  The SiPM signals are passed through a summing transimpedance pre-amplifier on the front end boards which has been designed to provide an output dynamic range of 2~V for signals from CsI(Tl) corresponding to energy deposits in the range of 50~keV to 3~MeV.  SensL also developed a custom motherboard to provide power distribution and fine bias control to all of the front-end boards.  The motherboard with the front end boards is shown in 
Fig.~\ref{fig:diagram}~d).  A benchtop power supply provides the motherboard with
$\pm 5$~V and -35.4~V for the pre-amplifier power and the SiPM bias respectively.
With all thirty-two SiPM channels powered and collecting data, the total current drawn is 750~mA.

The absorber layer consists of twenty-five NaI(Tl) crystals of dimension 2.5 x 2.5 x 4~cm$^3$.  Each is read out separately by a 
Hamamatsu\footnote{Hamamatsu Photonics K.K., Hamamatsu City, Japan} R1924A PMT.  The NaI(Tl) plus PMT assemblies were custom designed for this application by Proteus\footnote{Proteus Inc, Chagrin Falls, OH, USA}, in order to minimize the dead space between the individual NaI(Tl) crystals.  The outer dimension of the NaI(Tl) plus PMT assemblies is 
3.2 x 3.2 x 15.2~cm$^3$ and they are packed into a five by five array of outer dimension 16 x 16 x 15.2~cm$^3$.  High voltage for the PMTs is supplied using a
 CAEN\footnote{CAEN S.p.A., Viareggio, Italy} HV 1535SN unit and the PMT high voltages are set such that each channel reconstructs the 662~keV photopeak of $^{137}$Cs at the same pulse height.

\section{Data Acquisition}
\label{sec:acquisition}

Signals from both the SiPM pre-amps and the PMTs are processed by 
CAEN V1740 digitizers.  The digitizers are used to trigger on the leading edge of the PMT pulses.  When any PMT pulse causes a trigger, the waveforms from all scatter and absorber channels are transferred to memory on a Linux workstation.  Unfortunately, with the CAEN V1740 in its current configuration, the trigger threshold must be common across a group eight channels, though there are variations in the channel pedestals of up to $\sim$20~keV.  This leads to an effective spread in the trigger threshold across channels.  The 
average PMT trigger threshold corresponds to a pulse energy of approximately 100~keV for runs which were taken with the $^{137}$Cs and $^{113}$Sn sources.  The runs taken with the $^{22}$Na source used a higher trigger threshold.

Integration of the waveform is performed on all channels once an event
is received at the workstation, and the resulting charges
are written to disk for further analysis off line.
Integrated charges are calculated from the samples falling within a fixed time
window relative to the trigger time, using the mean of the samples
prior to the trigger for baseline subtraction.
For the scatter detector, the energy scale is parametrized linearly using the 662~keV line of 
$^{137}$Cs and the zero-energy pedestal.
For the absorber detector, the energy scale is parametrized with a third-order polynomial using 
emissions over the range 40~keV to 2614~keV from $^{241}$Am, $^{113}$Sn, 
$^{152}$Eu,
$^{22}$Na, $^{137}$Cs, $^{40}$K and $^{208}$Tl.

Table~\ref{tab:runs} lists the runs which were taken for this study, showing the experimental configuration and the number of triggered events, $N_{\mbox{\scriptsize trig}}$, for each run.  
Four runs were taken with the $^{137}$Cs source positioned along the x-axis (azimuthal angle  $\phi \sim 0$) at different values of $\theta$, the polar angle between the z-axis and the line between the origin and the source.  
One run was taken with a $^{113}$Sn source, to investigate the behaviour of the imager at lower energies.
Two runs were taken with a $^{22}$Na source using the 1274~keV photopeak.  
In one of the $^{22}$Na runs, run G, the rear scatter layer was removed so that only one scatter layer remained.  This was done to investigate the benefit of having more scatter material for higher energy sources. 
\begin{table}[!t]
\centering
\begin{tabular}{|c|c|c|c|c|l|c|c|}
\hline
\multirow{3}{*}{Run}&\multicolumn{5}{|c|}{Source} &\multirow{3}{*}{Livetime}&\multirow{3}{*}{$N_{\mbox{\scriptsize trig}}$}\\
\cline{2-6}                                            
          &Id              &Energy & BR    &Emission                      &   $\theta$          &  &  \\
          &                &[keV]  & [\%]  &Rate [s$^{-1}$]               &  [$^\circ$]&  [s] &  \\
\hline                                                                     
A          & $^{137}$Cs     & 662   &  85  &  ($2.2 \pm 0.3$)$\times 10^7$ & 0.6      & 11~479     & 12$\times 10^6$         \\
\hline                                                                     
B          & $^{137}$Cs    & 662   &  85   &  (2.2 $\pm$ 0.3)$\times 10^7$ & 10.8     & 11~374     & 12$\times 10^6$         \\
\hline                                                                     
C          & $^{137}$Cs     & 662   &  85  &  (2.2 $\pm$ 0.3)$\times 10^7$ & 20.3     & 11~340     & 12$\times 10^6$          \\
\hline                                                                     
D          & $^{137}$Cs     & 662   &  85  &  (2.2 $\pm$ 0.3)$\times 10^7$ & 28.6     & 22~401     & 24$\times 10^6$          \\
\hline                                                                     
E          & $^{113}$Sn     & 392   &  64  &  (1.4 $\pm$ 0.2)$\times 10^7$ & 10.8     & 32~907     & 30$\times 10^6$          \\
\hline                                                                     
F          & $^{22}$Na      & 1274  &  100 &  (2.4 $\pm$ 0.3)$\times 10^7$ & 10.8     & 26~418     & 24$\times 10^6$          \\
\hline                                                                     
G          & $^{22}$Na      & 1274  &  100 &  (2.4 $\pm$ 0.3)$\times 10^7$ & 10.8     & 8~511      & 8$\times 10^6$           \\
\hline
\end{tabular}
\caption{\label{tab:runs}
The run configurations are defined including the photopeak energy under study, its branching ratio (BR) and the source emission rate.
The number of events triggered is listed for each run.  Note that for run G only the forward scatter layer is present.}
\end{table}
Source positions were measured to 1~mm uncertainty using a 
Nikon Nivo$^{5.M}$ total station.
All runs were taken with the source at approximately the same distance $R$ from the origin; $R$ varies from 760.9~cm to 761.2~cm.
All runs were taken with the source in an approximately horizontal plane, with the y position of the
source varying from -7.0~cm to -8.2~cm.
We will use run B with the $^{137}$Cs source at $\theta = 10.8^\circ$ as the canonical setup.
The analysis will be described for that run in full while the results for the other runs will be presented in a briefer format.

\section{Event Selection}
\label{sec:selection}

\begin{figure}
   \begin{center}
   \begin{tabular}{c}
   \begin{overpic}[width=7.2cm]{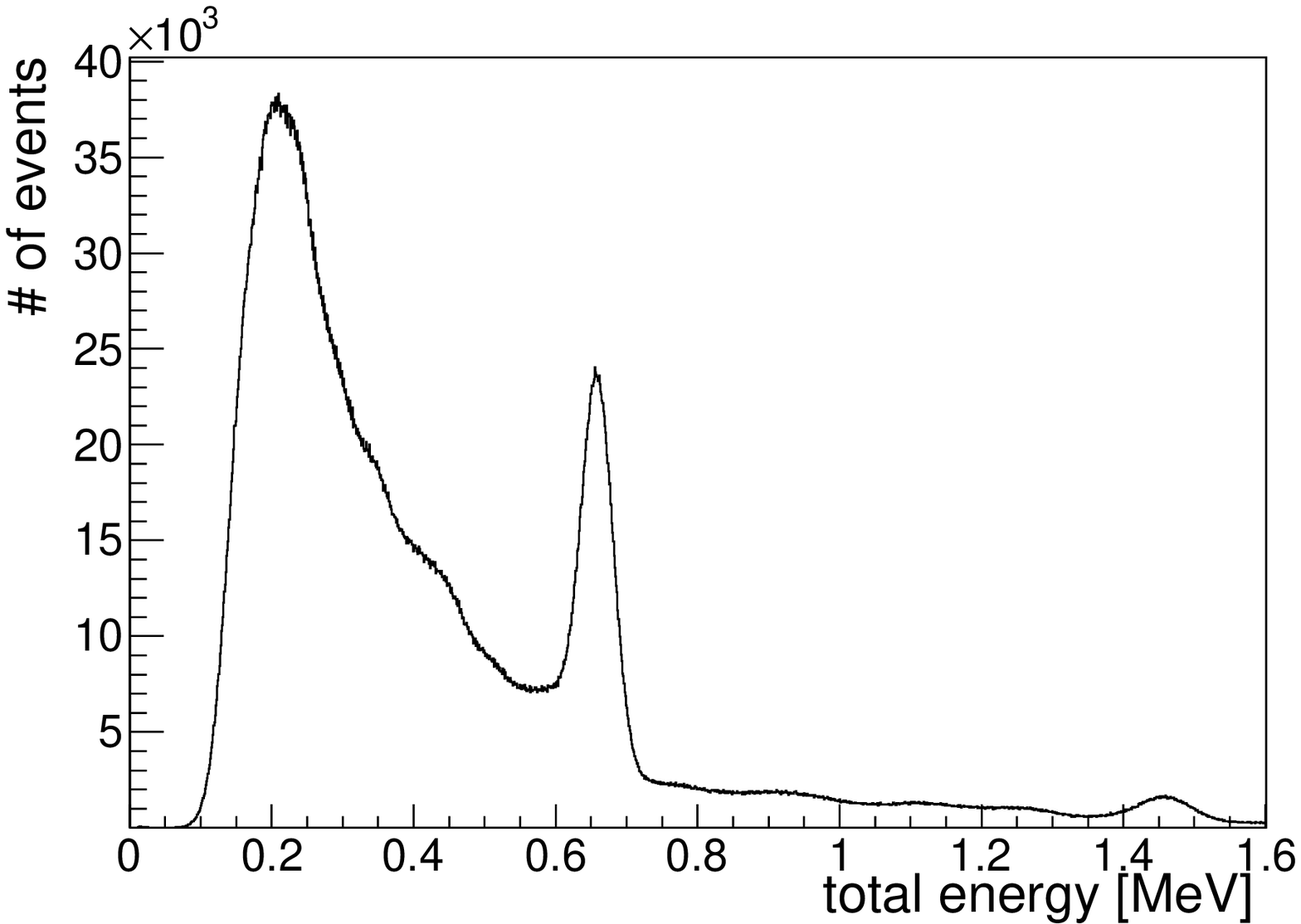}\put(80,53){a)}\end{overpic}
   \begin{overpic}[width=7.2cm]{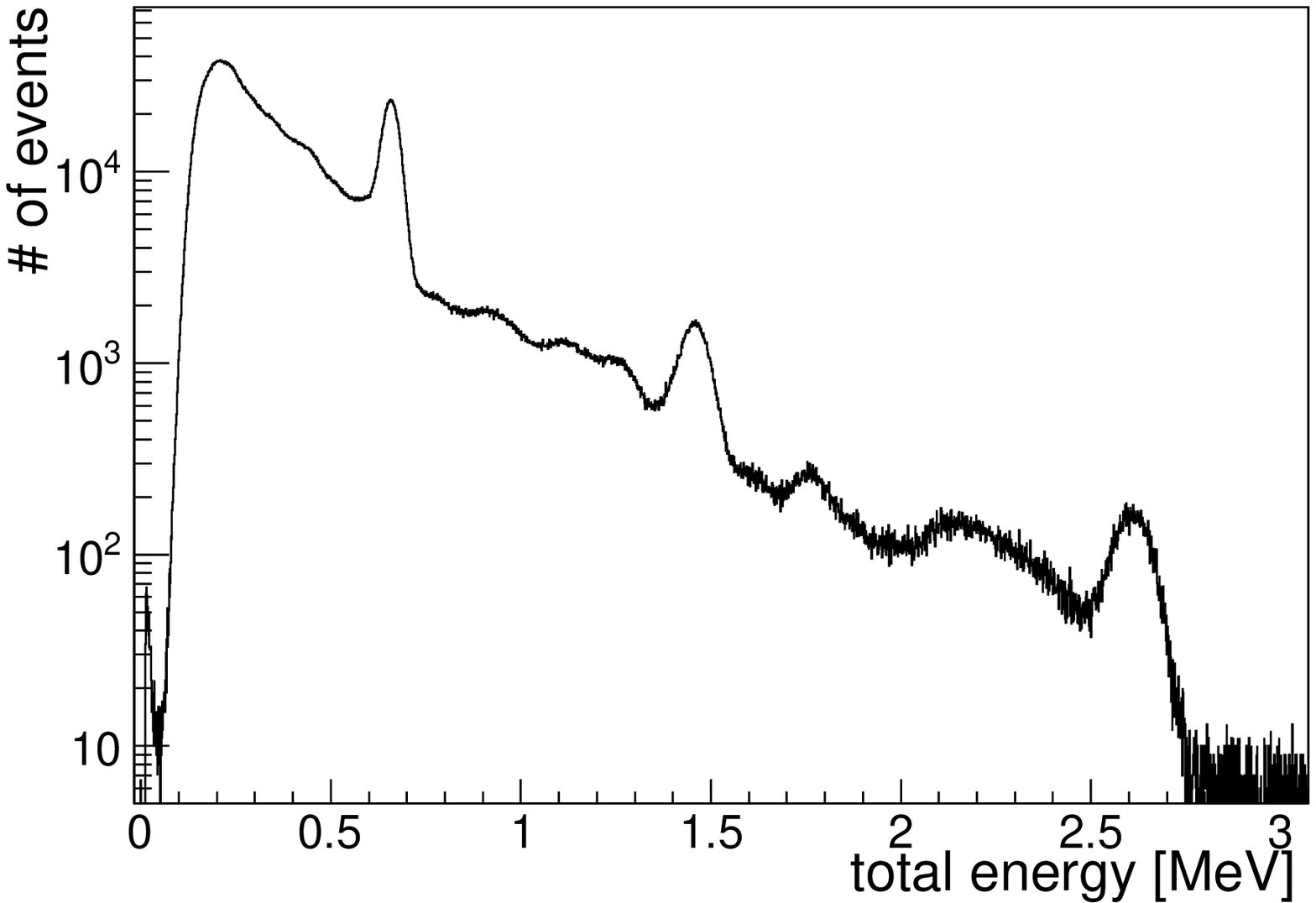}\put(80,53){b)}\end{overpic}\\
   \begin{overpic}[width=7.2cm]{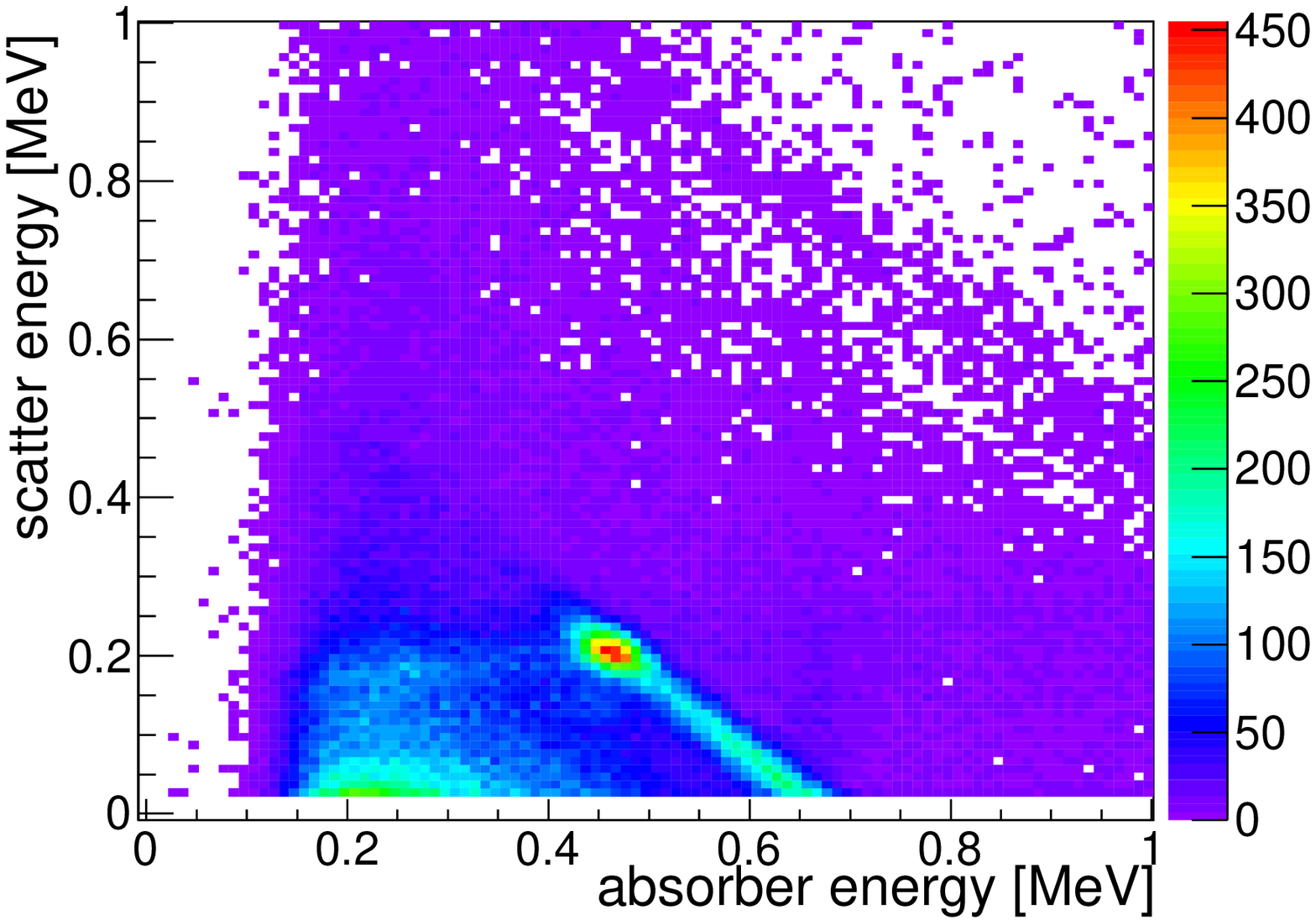}\put(13,51){c)}\end{overpic}\hspace{7.2cm}
   \end{tabular}
   \end{center}
   \caption
   { \label{fig:spectrum} 
Run B, $^{137}$Cs source at $\theta = 10.8^\circ$. The spectrum of the sum of energies deposited in the scatter and absorber detectors, where energy deposits $> 20$~keV ($> 25$~keV) in the absorber (scatterer) have been included in the sum is shown a) in linear scale and b) in logarithmic scale.  c) Scatter energy versus absorber energy, for all triggered events in which there is exactly one energy deposit in one of the scatter layers and one energy deposit in the absorber.}
   \end{figure}

The total energy spectrum for run B, for all triggered events, including only those 
energy deposits in the absorber with energy greater than 20~keV, 
and those energy deposits in the scatter layers with energy greater than 25~keV,
is presented in Fig.~\ref{fig:spectrum}~a).  
An energy resolution of 7.6\%~FWHM at 662 keV is obtained.  
Thus, the detector has sufficient resolution to be used as a spectrometer to identify radioactive contaminants quickly.  
Note that the total energy spectrum is dominated by events in which most or all of the energy has been deposited in the absorber layer.  
The energy resolution at 662~keV obtained from the CsI(Tl)+SiPM channels alone is on average 8.0\%~FWHM with a spread of 0.5\%~FWHM over the thirty-two channels.  
The NaI(Tl)+PMT channels give 7.4\%~FWHM with a spread of 0.2\%~FWHM over the twenty-five channels.  

Fig.~\ref{fig:spectrum}~b) shows the same data as in 
Fig.~\ref{fig:spectrum}~a) but this time extended to higher energies
and with a logarithmic y-scale.
Peaks from naturally 
occurring $^{40}$K (1460~keV), $^{238}$U
(1764~keV from $^{214}$Bi daughter) and $^{232}$Th (2614~keV from $^{208}$Tl daughter) are clearly visible.
This demonstrates the sensitivity of the instrument.
It can be used in situations where either weak source strength or heavy shielding reduces the signal to near the levels of naturally occurring background. 

The scatter energy  ($\Escat$) for run B is shown in Figure~\ref{fig:spectrum} c) versus the absorber energy ($\Eabs$), for all events with exactly one energy deposit in either of the 
two scatter layers and one energy deposit in the absorber layer.  We can see that the trigger threshold corresponded to an energy deposit of around 100~keV in the absorber channels.  
 The $^{137}$Cs photopeak is clearly apparent as the diagonal line with 
$\Escat + \Eabs \sim 662$~keV.
Note also the strong peak in the number of events at $\Escat \sim 200$~keV and
$\Eabs \sim 460$~keV.  These are ``backscatter'' events which scatter first in the absorber detector and subsequently interact in the scatter detector.  Under the assumption that the scatter detector energy is the first energy deposit, these events will lead to a mis-reconstruction of the Compton scatter angle and smear the reconstructed image.  Fortunately, for each photopeak energy the backscatter events are tightly peaked at $\Eabs$ and $\Escat$ energies which can be calculated from Compton kinematics.  They are thus easily rejected from the image reconstruction. 

The good forward-going Compton coincidence events lie in the photopeak and have scatter energy deposits extending to very low energies.
There is also noise in the scatter channels coming from the SiPMs, leading to a class of events with 
$\Escat$ extending from 0 to around 25~keV.  The coincidence energy threshold in the scatter detector is therefore set at 30~keV in order to collect
as many of the good events as possible, while rejecting noise.

To reconstruct the source position using our imager, we define the coincidence event selection to include only those events with exactly one hit of energy $\Escat > 30$~keV in either of the two scatter planes, and exactly one hit of energy $\Eabs > 20$~keV in the absorber layer.  Events are required to have $\Escat +\Eabs$ within roughly two sigma of the photopeak energy,
specifically 35~keV, 40~keV, and 80~keV of the photopeak energies 
for $^{113}$Sn, $^{137}$Cs and $^{22}$Na, respectively.

Backscatter rejection is applied differently for each isotope.  For $^{113}$Sn and
$^{137}$Cs, events are accepted for $\Escat < 120$~keV and
$\Escat < 170$~keV respectively.
For $^{22}$Na, events which satisfy either $\Escat < 170$~keV
or 280~keV $< \Escat < 635$~keV are accepted.

\section{Results}
\label{sec:results}

\subsection{Efficiency and angular resolution}

We define efficiency as the number of events which pass the coincidence selection, 
minus the number of natural background events derived from a run without sources present,
divided by the total number of gamma rays which cross the 12~x~12~cm$^2$ area of the first scatter layer during the run.  The uncertainty on the strength of the source dominates all other uncertainties in the measurement of efficiency.  
The efficiency measurements for all runs are presented in Table~\ref{tab:results}.
These measurements may be used to compare the effective area of our design with the effective areas of other Compton imagers under development.

The Angular Resolution Measure (ARM) is defined as $\thetaCR - \thetageom$, where
$\thetaCR$ is the Compton scatter angle as it is reconstructed from the measured energy deposits and $\thetageom$ is the actual angle between the line which joins the two energy deposits and the line from origin to source.
The ARM shows how well the detector reconstructs the source position, and together with
efficiency, ARM can be used to compare the performance of various Compton imager designs in a quantitative manner.

\begin{figure}
   \begin{center}
   \begin{tabular}{c}
   \begin{overpic}[width=7cm]{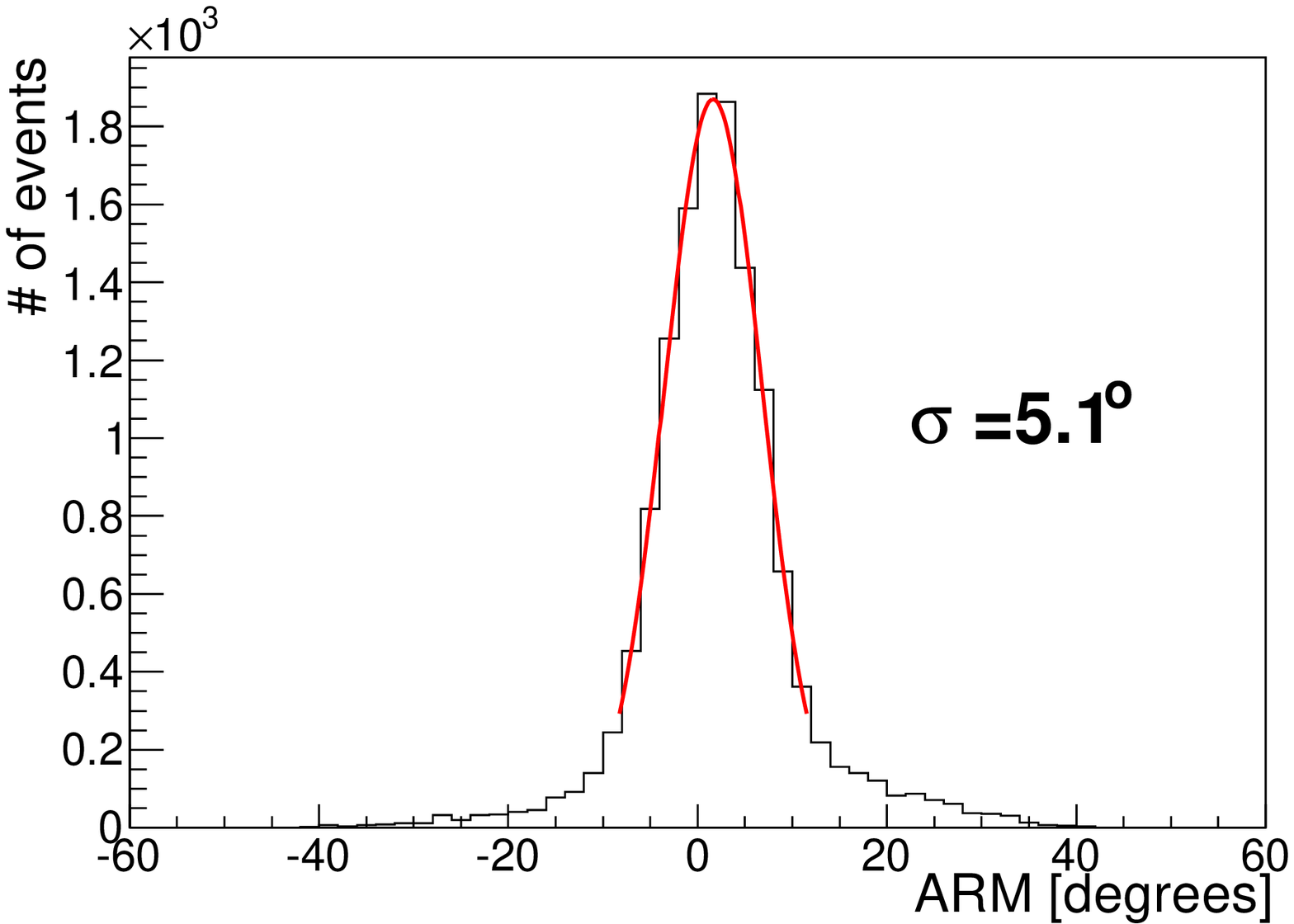}\put(15,54){a)}\end{overpic}
   \begin{overpic}[width=7cm]{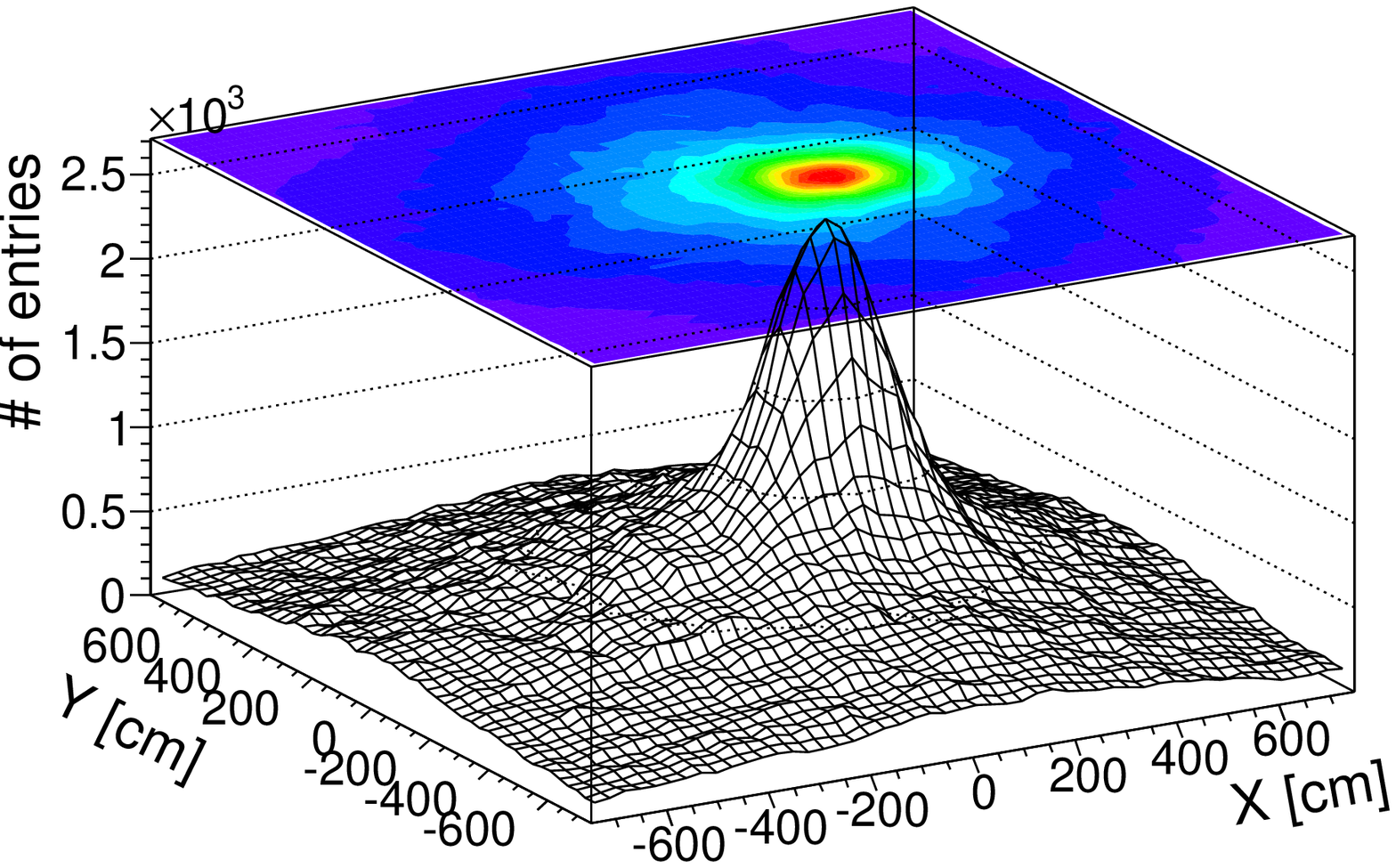}\put(80,58){b)}\end{overpic}
   \end{tabular}
   \end{center}
   \caption
   { \label{fig:backprojection} 
Run B, $^{137}$Cs source at $\theta = 10.8^\circ$. a) ARM distribution and b) backprojection image, for all coincidence events.}
   \end{figure}

The ARM distribution for coincidence events, for run B, is shown in 
Fig.~\ref{fig:backprojection}~a).  The event selection is dominated by forward going scatters which are well reconstructed; the ARM distribution exhibits a peak at zero with a width of $\sigmaARM = 5.1^\circ$.  There is also a baseline of poorly reconstructed events which are due to the remaining naturally occurring background, mis-assigned backscatters and multiple scatter events.  
The ARM distributions for the other runs look similar.  The widths of Gaussian curves fit to the peaks of the ARM distributions are presented in Table~\ref{tab:results} for all runs.

\subsection{Imaging -- $^{137}$Cs}

\subsubsection{Backprojection image}

By projecting the Compton cones from all of the events in the coincidence selection onto an image plane, a backprojection image may be obtained.
To present the image in x and y coordinates, the known distance of the source in z was used.  Note that this known position is in no way required to locate 
the source in $\theta$ and $\phi$.
(In future work with the larger detector a determination of the z position of the source may be possible in some scenarios.)
A backprojection image for run B is shown in Fig.~\ref{fig:backprojection}~b).  The backprojection image shows that the imager can function to localize a source of radiation in its field of view, without extensive treatment of the data.

\subsubsection{Image reconstruction}

To further quantify the imaging performance of the imager, we have applied a $\chi^2$ minimization algorithm to the backprojected Compton cones to find the common point of closest approach~\cite{us_2009}.
Advantages of the reconstruction algorithm are its ability to reject
background and its use of the differing uncertainties on each
backprojected cone,  thus permitting a measurement of the source
position with high precision and accuracy.

\begin{figure}
   \begin{center}
   \begin{tabular}{c}
   \hspace{1cm}\begin{overpic}[height=4.5cm]{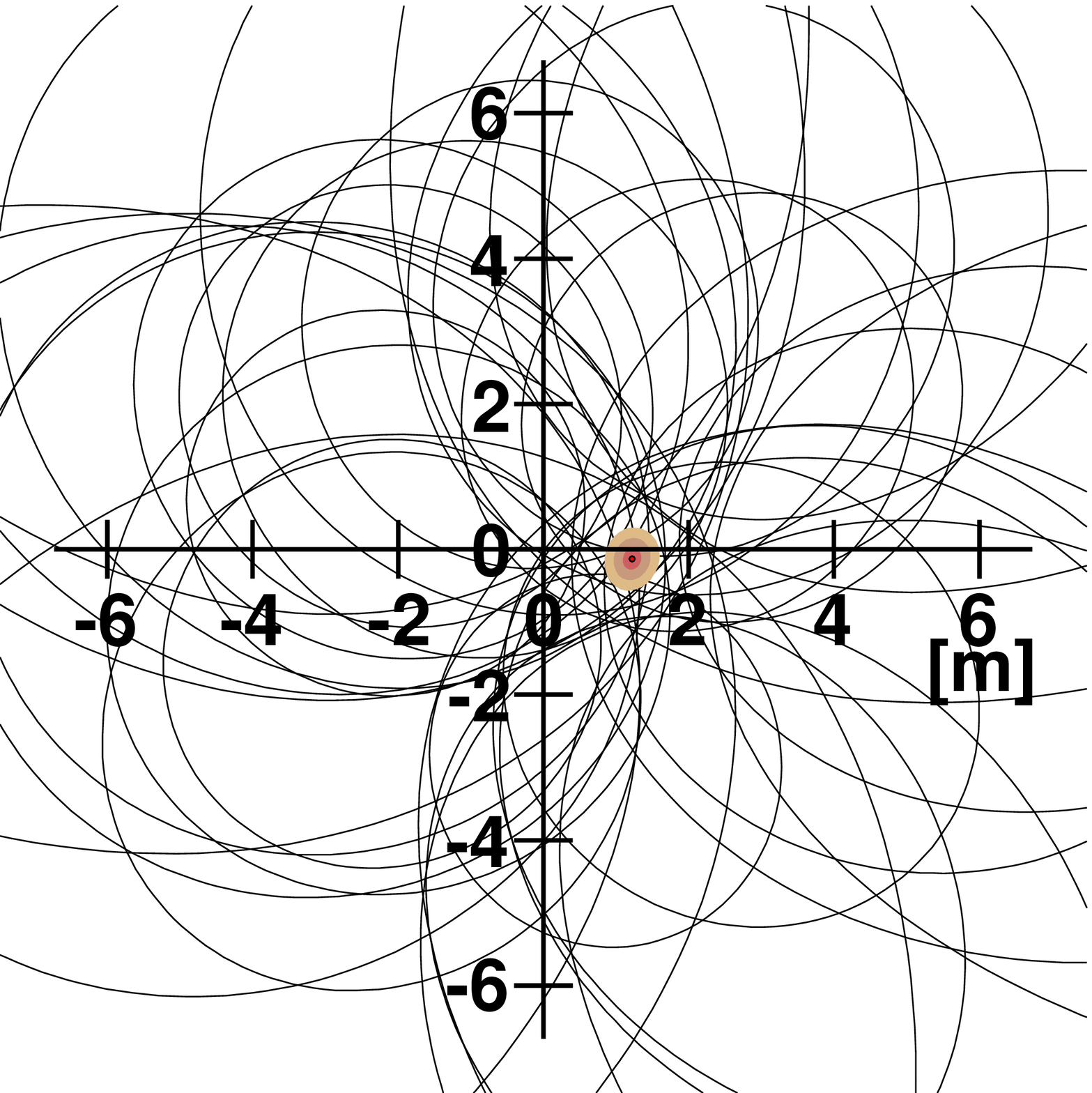}\put(-15,82){a)}\end{overpic}
   \hspace{1.5cm}\begin{overpic}[width=7cm]{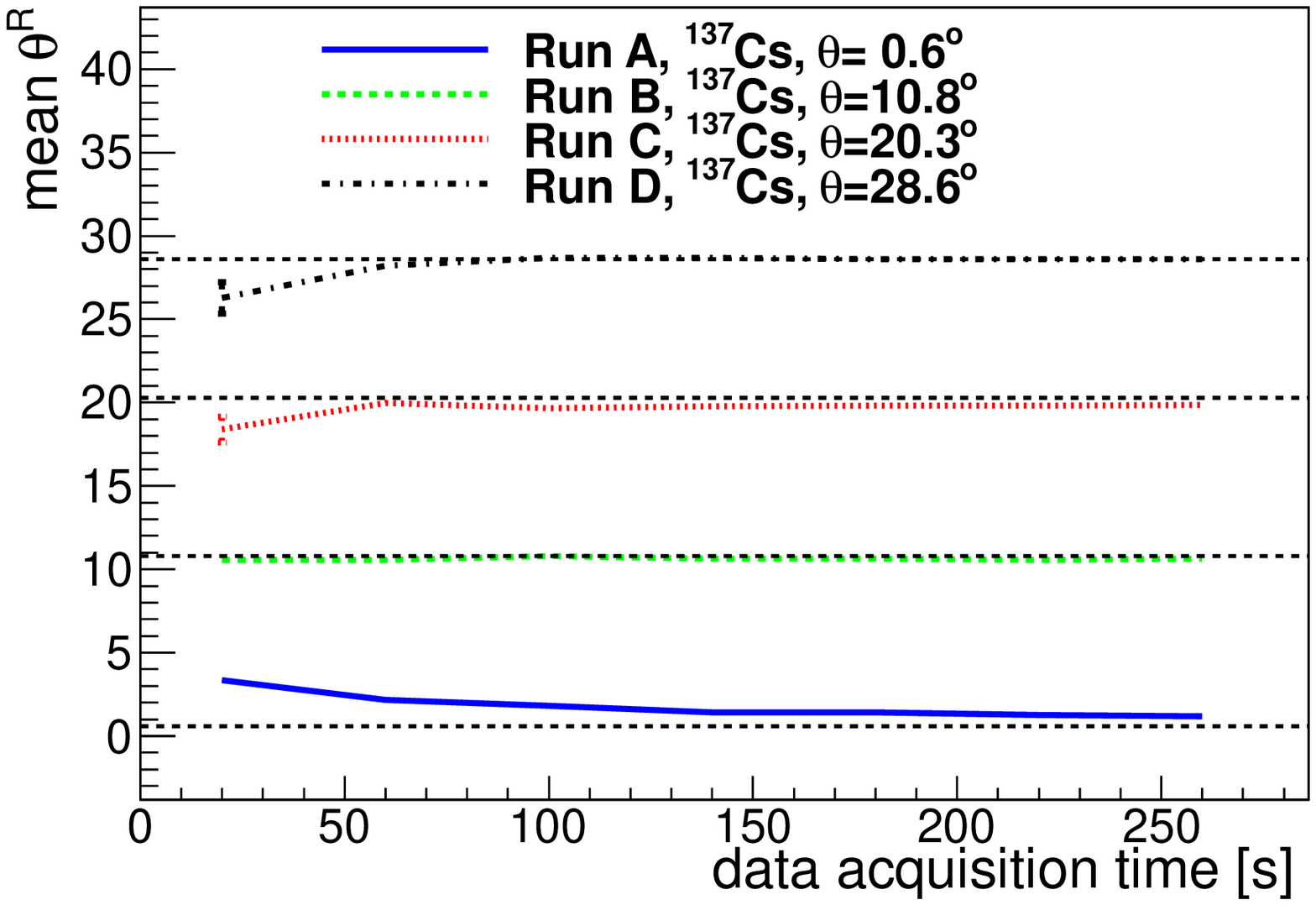}\put(80,52){b)}\end{overpic}\\
   \begin{overpic}[width=7cm]{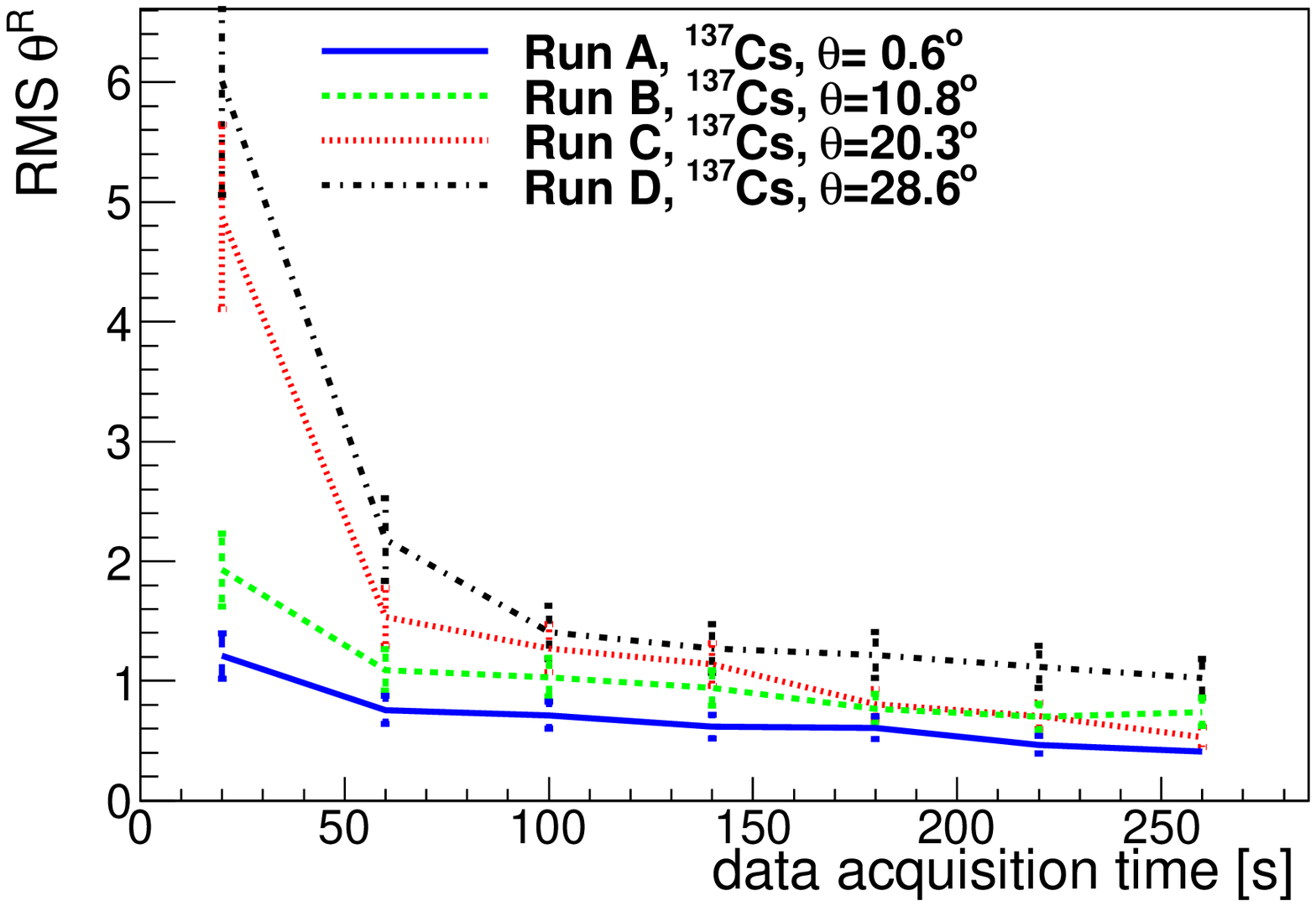}\put(80,52){c)}\end{overpic}
   \hspace{.2cm}\begin{overpic}[width=7cm]{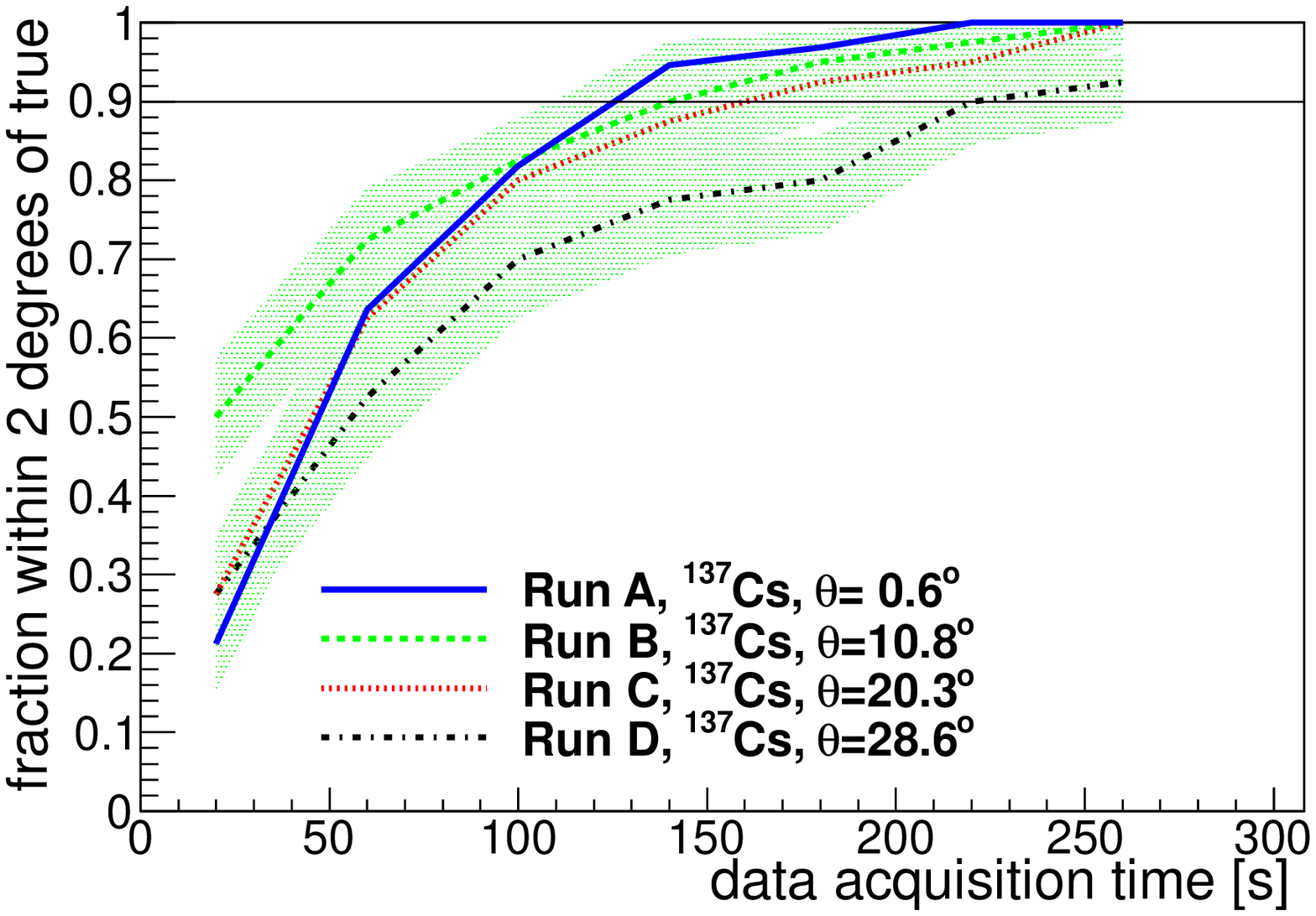}\put(80,46){d)}\end{overpic}
   \end{tabular}
   \end{center}
   \caption
   { \label{fig:time_to_image} 
a) A reconstructed image for one 20~s trial, with shaded areas indicating the 
1~$\sigma$, 2~$\sigma$ and 3~$\sigma$ contours for run B, $^{137}$Cs source at $\theta = 10.8^\circ$.
b) Reconstruction accuracy and c) precision as a function of trial acquisition time for runs A, B, C and D. d) The fraction of time that the 
source is reconstructed to within 2$^\circ$ of the correct source position, as a function
of trial acquisition time, for runs A, B, C and D.}
   \end{figure}

The result of the image reconstruction for just 20~s of data (53 events) from run B, is shown in Fig.~\ref{fig:time_to_image}~a).  The
one, two and three sigma contours are indicated by the shaded areas.  
The source
position reconstructs to $\thetaR=(9.3 \pm 0.9)^\circ$  and 
$\phiR=(-6 \pm 7)^\circ$ at one standard deviation.

To understand how long it takes the imager to produce an image of a certain quality, a number of trials (32 or 40), each of a certain acquisition time, is selected from each data set.  Then we can look, as in Fig.~\ref{fig:time_to_image}~b) at the mean of the reconstructed source position, $\thetaR$, over several trials, as a function
of acquisition time.  We find that, aside from the run B data when the source was positioned at 
10$^\circ$ off-axis, the accuracy improves with increasing acquisition time.  (Note that the systematic loss of accuracy for low acquisition times is well reproduced in Monte Carlo simulation and will be corrected for in future work.)
We can also look at how the spread of the measured source positions improves with acquisition time.
Fig.~\ref{fig:time_to_image}~c) shows the root mean square of the distribution of $\thetaR$ 
versus acquisition time, for runs A through D.
We find that the precision is worse for off-axis sources, and improves 
strongly with acquisition time, reaching $\sim 2^\circ$ by $\sim 70$~s of
acquisition time, across the field of view.
(Note that the full detector in preparation has been designed for an improved field of view.)

Fig.~\ref{fig:time_to_image}~d) shows a plot which we use to define a measure of the ``time to image'' the source.  We look at the percentage of trials for which the source position is reconstructed correctly to within two degrees, as a function of acquisition time.  Time to image is defined as the acquisition time by which the source is 
reconstructed correctly to within two degrees, 90\% of the time.  Thus we see that for run A, with
the source positioned on axis, the time to image is just over two minutes.  The time to image is longer for sources which are further off-axis, as expected given the worsening of image precision for off-axis sources shown in Fig.~\ref{fig:time_to_image}~c).
The times to image for the various runs are presented in Table~\ref{tab:results}.

For ease of comparison of these results with those of the other sources, we have
also scaled the times to image to obtain effective times to image for a hypothetical bare source of strength 1~mCi and 
branching ratio equal to 100\%.  Note that in doing this, we have neglected the effect on the time to image of any naturally occuring background contamination present in the final fit of the reconstruction algorithm, estimated to be
at most 5\% of the events for Cs-137.

\subsection{Imaging -- $^{113}$Sn}

\begin{figure}
   \begin{center}
   \begin{tabular}{c}
     \begin{overpic}[width=7cm]{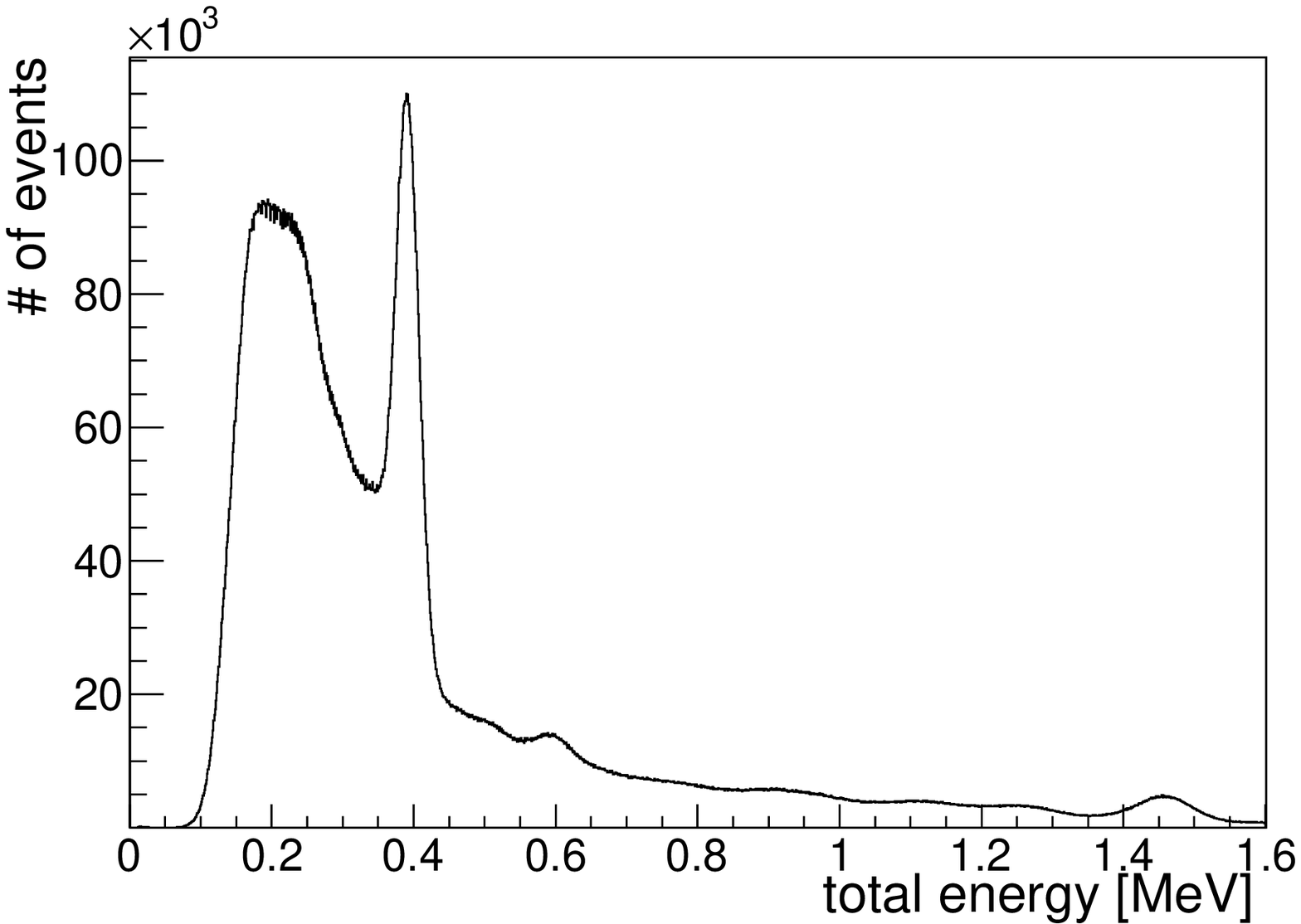}\put(80,53){a)}\end{overpic}
     \hspace{.2cm}\begin{overpic}[width=7cm]{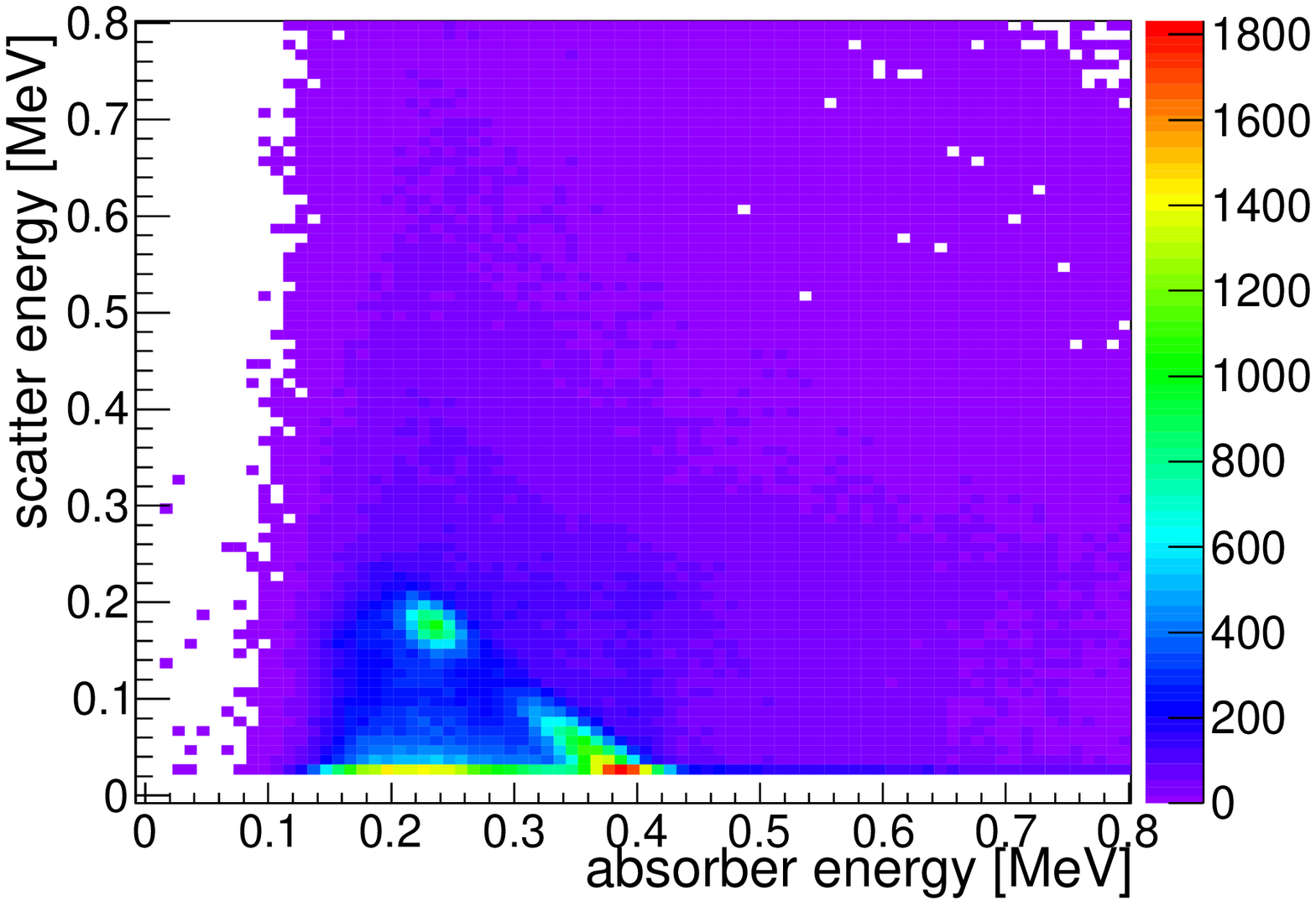}\put(13,53){b)}\end{overpic}\\
     \begin{overpic}[width=7cm]{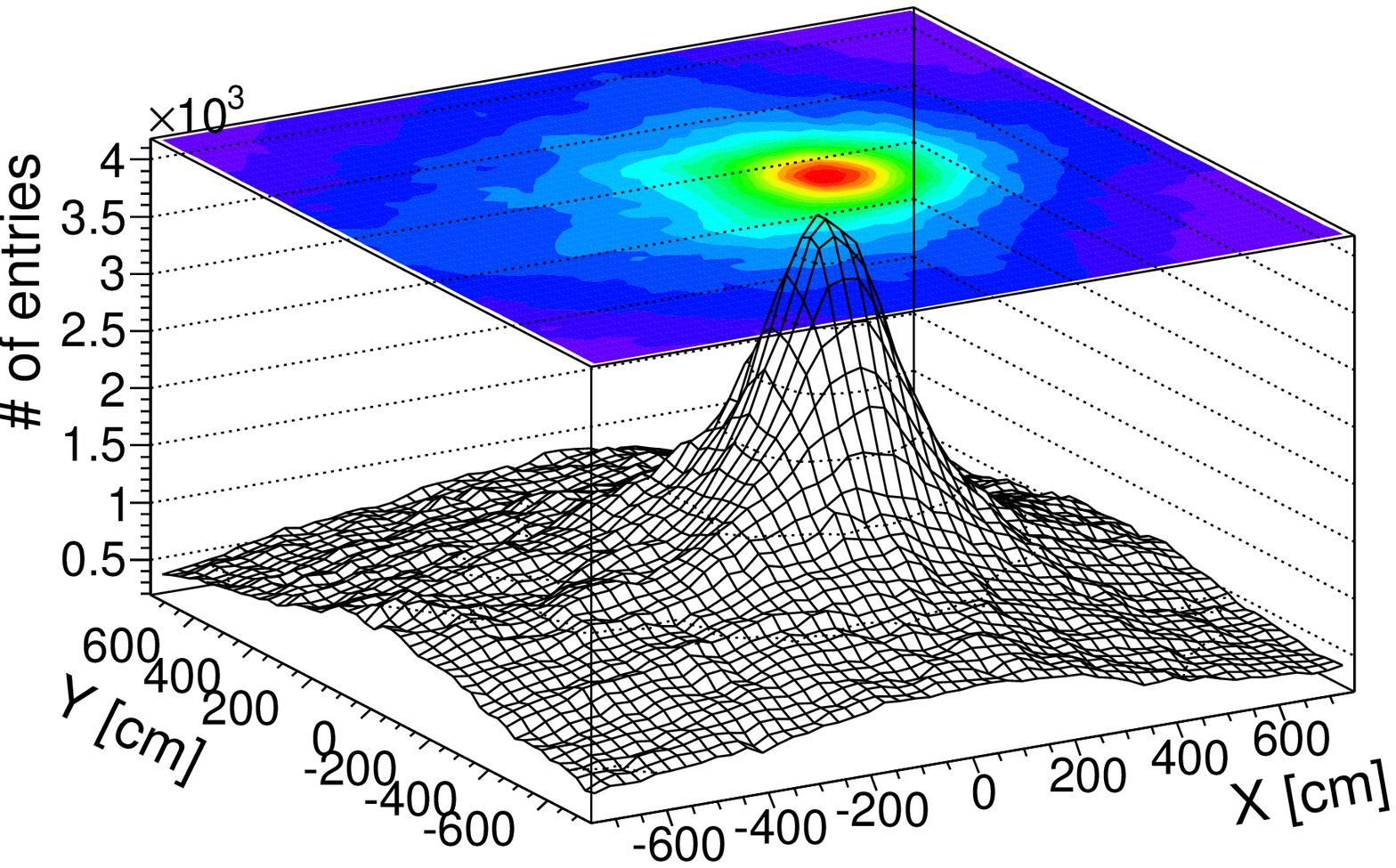}\put(80,58){c)}\end{overpic}
     \hspace{.2cm}\begin{overpic}[width=7cm]{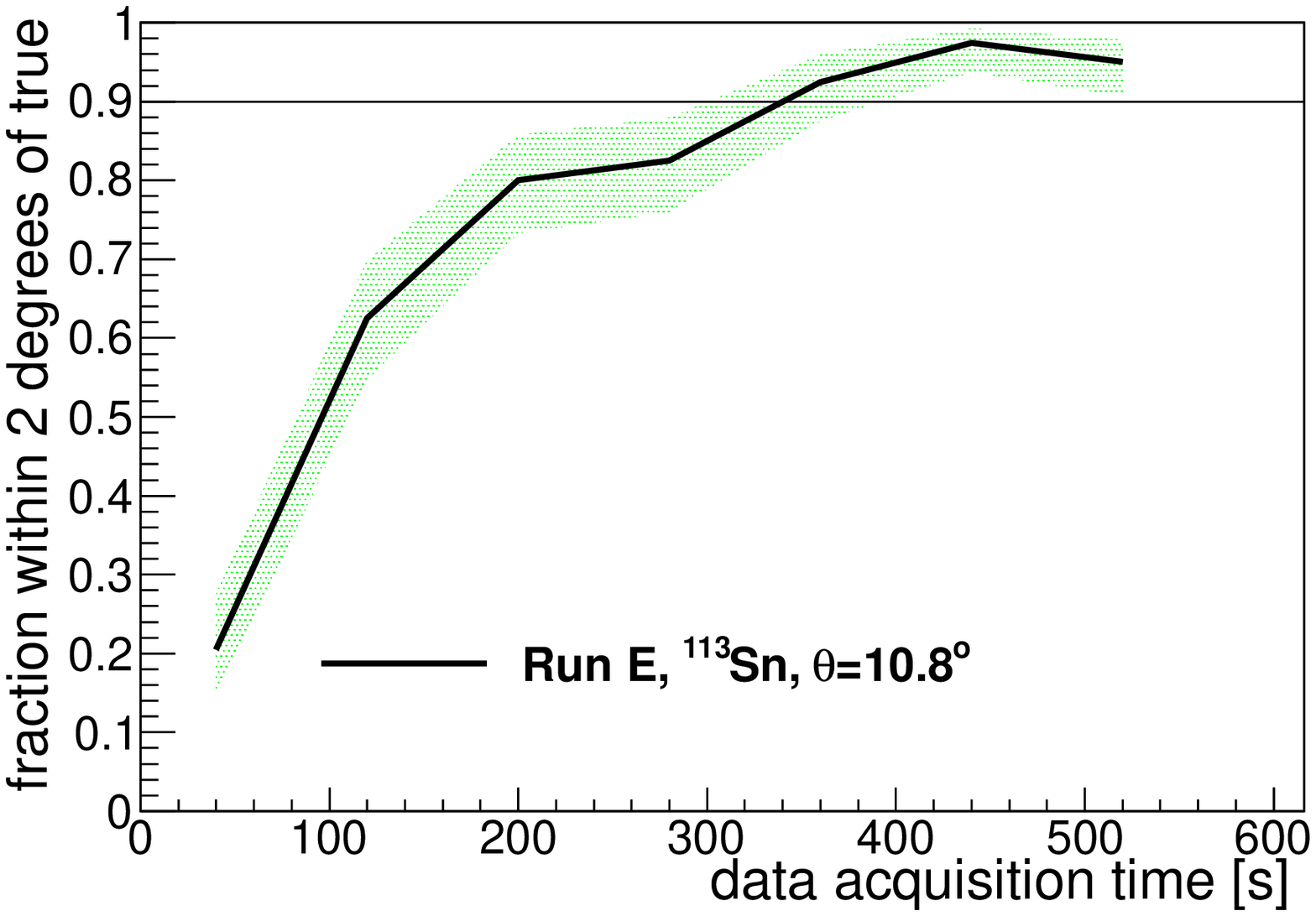}\put(78,46){d)}\end{overpic}
   \end{tabular}
   \end{center}
   \caption
   { \label{fig:Sn113} 
Run E, $^{113}$Sn source at $\theta = 10.8^\circ$. a) The energy spectrum, including all events which pass the trigger, and all energy deposits over 20~keV (25~keV) in the absorber (scatterer). 
b) The scatter energy versus the absorber energy for all events with some energy deposited in either scatter plane and some energy deposited in the absorber.
c) A backprojection image for events passing the coincidence selection.
d) The fraction of trials reconstructing the source location correctly to within two degrees versus acquisition time. }
   \end{figure}

Results for the $^{113}$Sn source, run E, are shown in Fig.~\ref{fig:Sn113}.  This source has a photopeak at 392~keV which is well reconstructed by the imager, with an energy resolution of 9.3\%~FWHM at 392~keV, as shown in Fig.~\ref{fig:Sn113}~a).  The forward going Compton scatters from this source deposit energy very close to the noise level in the scatter detector, see Fig.~\ref{fig:Sn113}~b).  Nevertheless the backprojection of the Compton cones for events passing the coincidence selection results in a clear image of the source location, as shown in Fig.~\ref{fig:Sn113}~c).  The time to image plot for the $^{113}$Sn source is presented in Fig.~\ref{fig:Sn113}~d).  
At ($340 \pm 40$)~s, the time to image is much longer than for the $^{137}$Cs source.  Even after correction for the difference
in source emission rates (neglecting natural background estimated at 9\%), 
the effective time to image, presented in Table~\ref{tab:results}, is significantly longer for the 
$^{113}$Sn source.  This is largely due to the poor ARM which results from the worsened energy resolution of the lower energy deposits.

\subsection{Imaging -- $^{22}$Na}

\begin{figure}
   \begin{center}
   \begin{tabular}{c}
   \begin{overpic}[width=7cm]{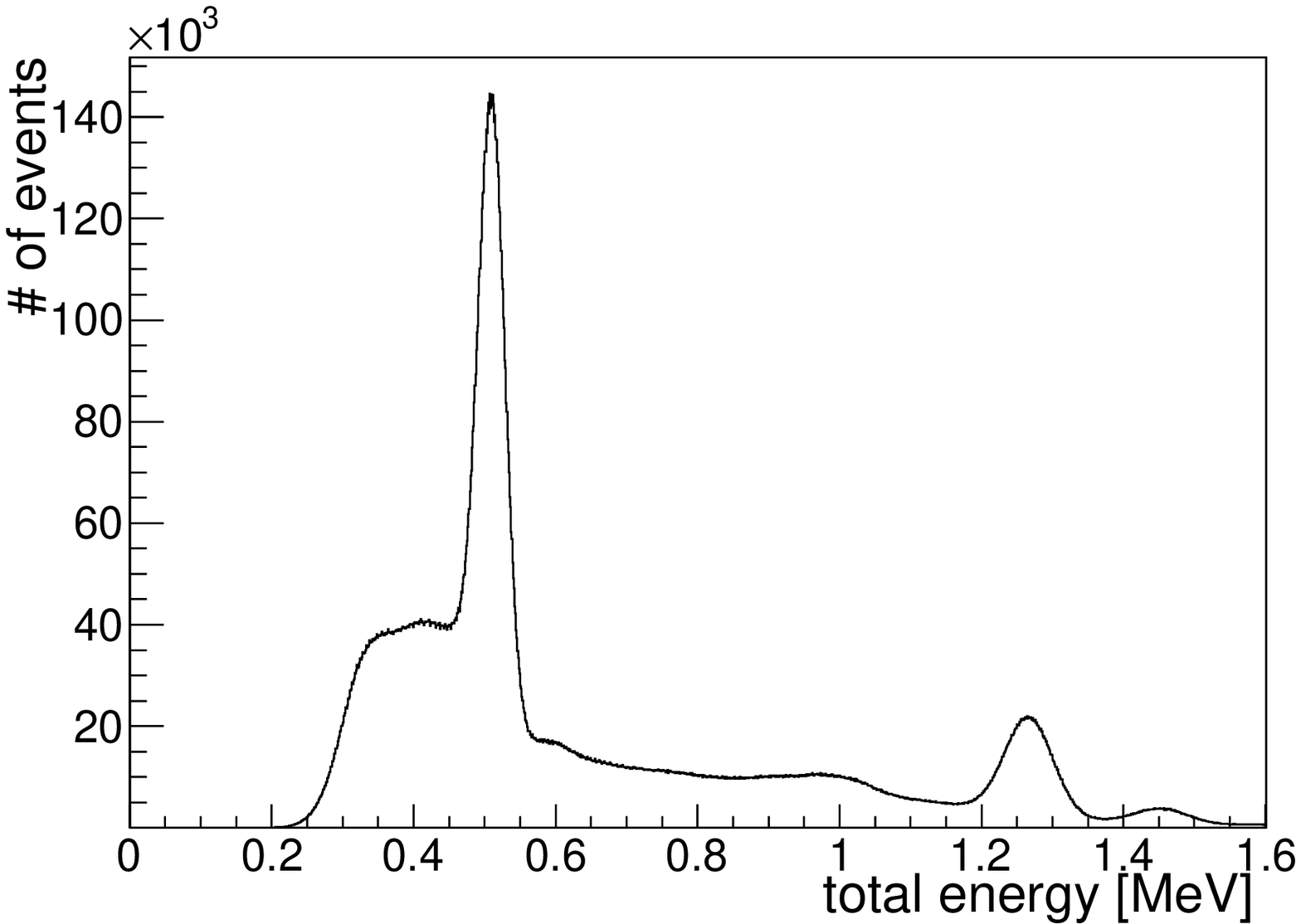}\put(80,53){a)}\end{overpic}
    \hspace{.2cm}\begin{overpic}[width=7cm]{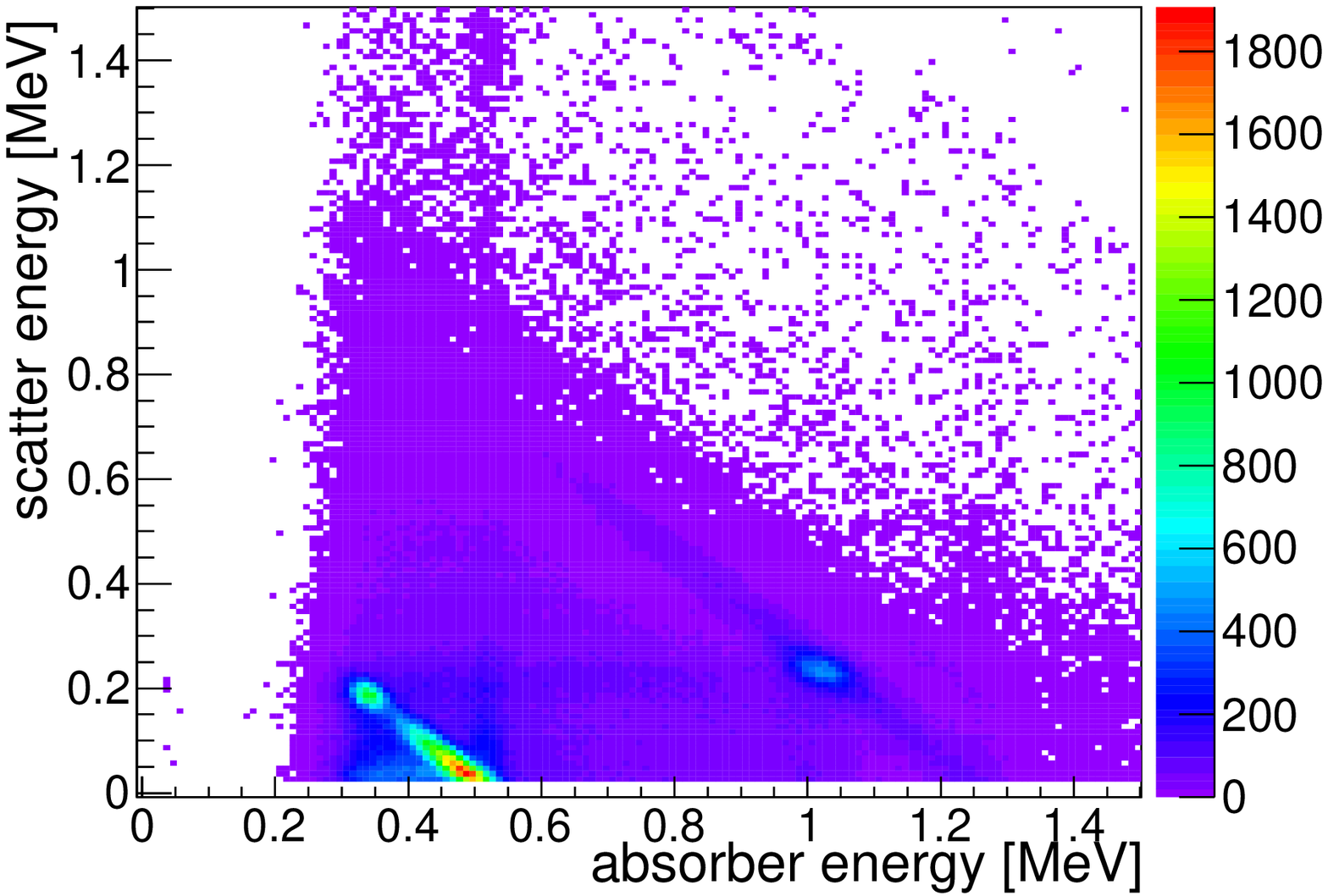}\put(13,53){b)}\end{overpic}\\
   \begin{overpic}[width=7cm]{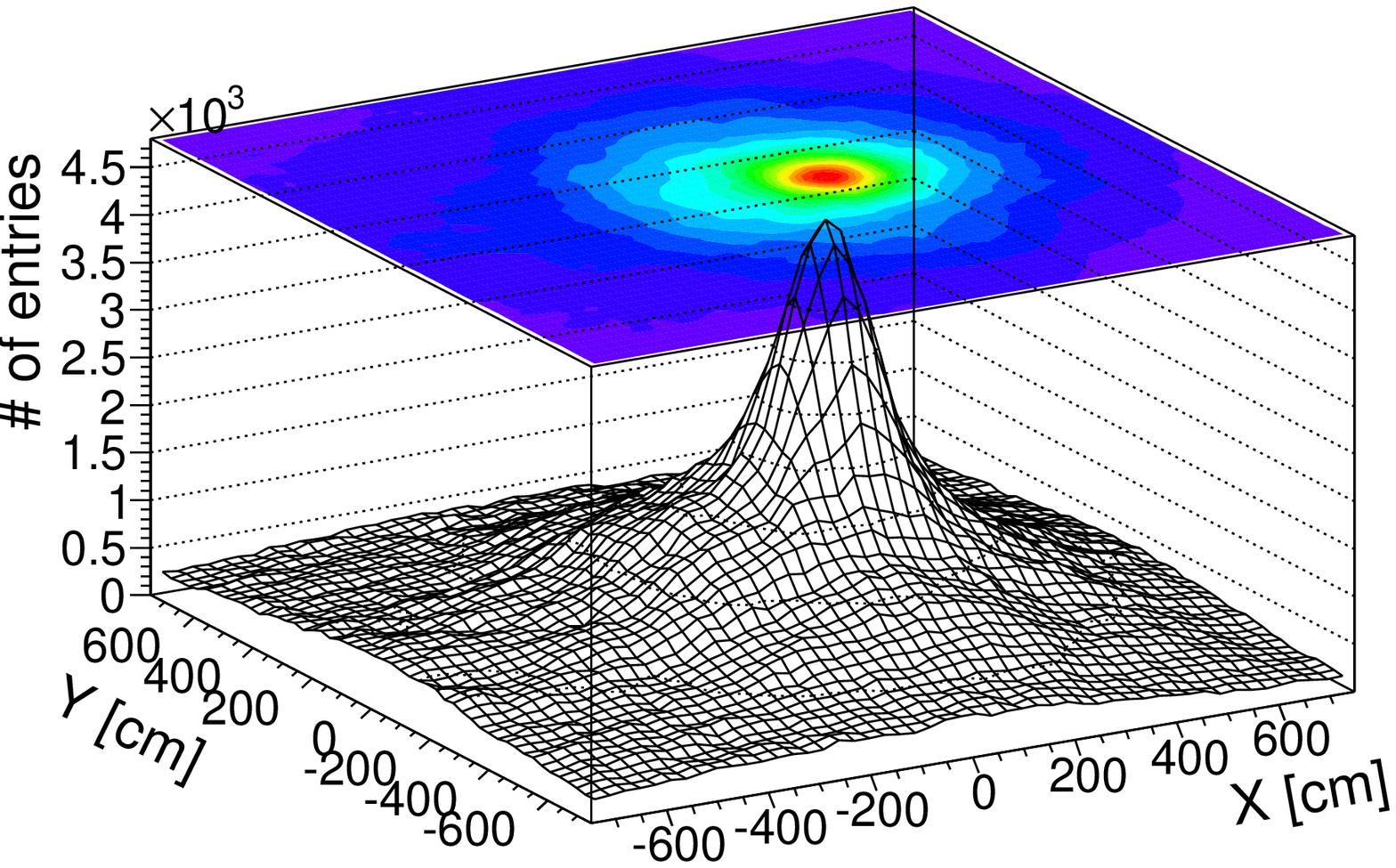}\put(80,58){c)}\end{overpic}
   \hspace{.2cm}\begin{overpic}[width=7cm]{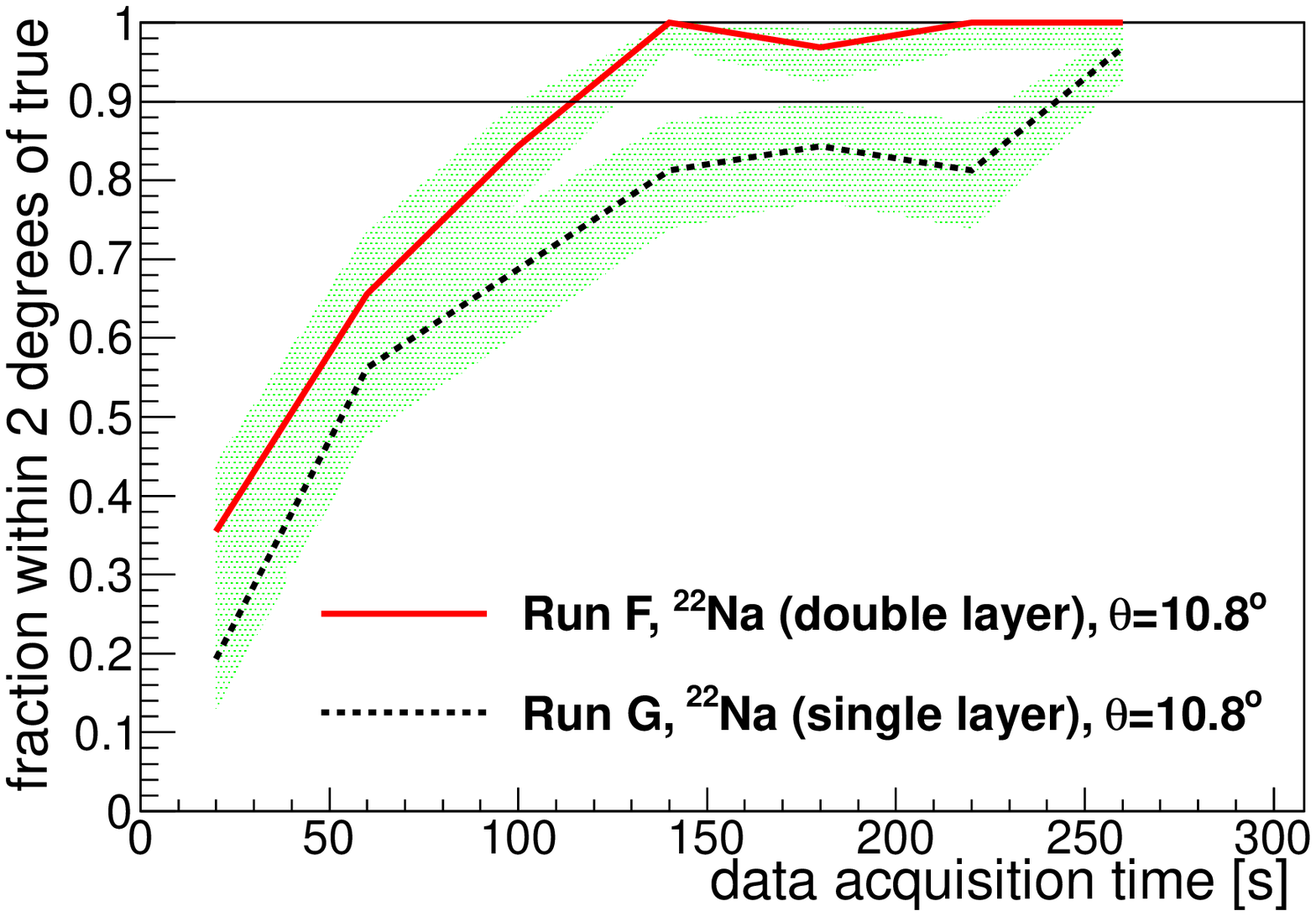}\put(78,46){d)}\end{overpic}
   \end{tabular}
   \end{center}
   \caption
   { \label{fig:Na22} 
Runs F and G, $^{22}$Na source at $\theta = 10.8^\circ$.
a) The energy spectrum, including all events which pass the trigger, and all energy deposits over 20~keV in the absorber and over 25~keV in the scatterer, for run F. 
b) The scatter energy versus the absorber energy for all events with some energy deposited in either scatter plane and some energy deposited in the absorber for run F.
c) A backprojection image for events satisfying the coincidence selection for run F.
d) The fraction of trials reconstructing the source location correctly to within two degrees versus acquisition time, for runs F and G. }
   \end{figure}

Results for the $^{22}$Na source, runs F and G, are shown in Fig.~\ref{fig:Na22}.
Both the 1274~keV gamma ray and the 511~keV annihilation peak peaks are well measured with 
the loose event selection as shown in
Fig.~\ref{fig:Na22}~a), with energy resolutions of
6.1\%~FWHM and 8.5\%~FWHM respectively.  In order to study the performance of our
detector at high energy, we considered only the 1274~keV photopeak, and raised the on-board trigger threshold for the absorber channels to $\sim 250$~keV.
Fig.~\ref{fig:Na22}~b) shows scatter energy versus absorber energy for all events depositing some amount of energy in both detectors for run F.  For the 1274~keV peak, the 
backscatter events deposit energy at $\Escat\sim$220~keV.
The backprojection image for events satisfying the coincidence selection 
demonstrates the ability of the system to localize the source very well, as shown in 
Fig.~\ref{fig:Na22}~c).  

Applying the image reconstruction algorithm, we find that the time to image for the 1274~keV peak is $(115 \pm15)$~s, as shown in Fig.~\ref{fig:Na22}~d).  The effective time to image for a 1~mCi source with a 1274~keV line with 100\% branching ratio would 
be $(75 \pm 15)$~s, where we have neglected an estimated background contribution
of 7\%.  (Note that although the efficiency appears to be much lower for $^{22}$Na, in fact a smaller proportion of the forward-going coincident events is rejected by the backscatter cuts, so it is not surprising to see no worsening of the time to image.)

For run G, we removed the rear scatter layer.  This produces an imager with the same overall external dimensions, but with less material for initiating the Compton scatter.  We see from the
time to image plot shown in Fig.~\ref{fig:Na22}~d) that without the second scatter layer, it takes longer for the imager to produce a good image.  This result was expected as the total amount of material in the scatter detector had been optimized by looking at the probability for a gamma ray to undergo one Compton scatter in the scatter material and then escape to the absorber detector~\cite{us_2009}.  Further work will examine other benefits of a multi-layer scatter detector including investigation of the class of events which undergo two Compton scatters before absorption.

\subsection{Summary}

\begin{table}[!t]
\centering
\begin{tabular}{|l|c|c|c|c|}
\hline
Run      &Efficiency      &$\sigmaARM$    &Time to image & Effective time to image\\

         &    [\%]         & [$^\circ$]    & [s]          & 1~mCi, BR=100\% [s]\\
\hline 
A        &0.52 $\pm$ 0.08  & 5.2 $\pm$ 0.1 & 125 $\pm$ 15 & 75 $\pm$ 15   \\
\hline     
B        & 0.56 $\pm$ 0.08 & 5.1 $\pm$ 0.1 & 140 $\pm$ 20 & 85 $\pm$ 20\\
\hline     
C        & 0.59 $\pm$ 0.09 & 5.3 $\pm$ 0.1 & 160 $\pm$ 25 & 95 $\pm$ 20 \\
\hline     
D        & 0.61 $\pm$ 0.09 & 5.3 $\pm$ 0.1 & 220 $\pm$ 30 & 130 $\pm$ 25 \\
\hline     
E        & 0.54 $\pm$ 0.08 & 6.2 $\pm$ 0.2 & 340 $\pm$ 40 & 125 $\pm$ 25 \\
\hline     
F        & 0.39 $\pm$ 0.06 & 5.1 $\pm$ 0.1 & 115 $\pm$ 15 & 75 $\pm$ 15  \\
\hline     
G        & 0.16 $\pm$ 0.02 & 4.2 $\pm$ 0.1 & $>$220       & $>$145 \\
\hline
\end{tabular}
\caption{\label{tab:results}
The efficiency of the detector and the width of the angular resolution measure is shown for each of the runs.  
Also shown is the time to image as determined for the particular source under study, as well as the effective time
to image for a hypothetical 1~mCi source with 100\% branching ratio.}
\end{table}

A summary of all of the results for efficiency, $\sigmaARM$ and time to image is provided in Table~\ref{tab:results}.  Our design performs well for source
energies extending from 392~keV to 1274~keV.  We find some degradation in the imager performance with the current setup, as the source is moved toward the edge of the field of view.  We find that the configuration with two scatter layers performs better at high energy than a configuration with a single scatter layer.  The results from this work are being used as guidance in a longer project we have underway to build a larger, fully optimized Compton imager.

\section{Conclusions}

We have shown that a Compton gamma imager using inorganic scintillating material for gamma detection, read out with silicon photomultipliers, can perform very well.
The small form factor of the SiPMs allows for the design of a compact, potentially human
transportable, device.
The low density of the SiPMs means that they cause minimal inefficiency due to scattering 
in dead material and their small power usage renders them suitable
for applications in the field.
Ultimately, we have found that with an imager of overall exterior 
dimension $\sim$~16 x 16 x 27~cm$^3$, made out of materials which are inherently ruggedizable, 
we can achieve image resolutions of around two degrees within two minutes, for isotopes of strength 1~mCi situated $\sim 8$~m away, across a wide range of energies.
We are encouraged by the success of this study and will continue to
develop this design further into a full-scale human-transportable imager based on 
scintillators with SiPM readout, capable of 
localizing a 10~mCi point source 40~m away to within a few degrees in under a minute.
This device will be
useful in radiological investigations and in incident remediation.

\section*{Acknowledgements}

Funding for this project was provided through Canada's Chemical, Biological, Radiological-Nuclear and Explosives Research and Technology Initiative.  This report is NRCan/ESS Contribution 20110314.





\bibliographystyle{elsarticle-num}
\bibliography{GI_SiPM}







\end{document}